\documentclass[onecolumn,showpacs,showkeys,preprintnumbers,amsmath,amssymb,nofootinbib,12pt]{revtex4}
\usepackage{graphicx}
\usepackage{dcolumn}
\usepackage{bm}
\usepackage{color}
\usepackage{graphics} 
\usepackage{graphicx}
\usepackage{epsfig}
\usepackage{epstopdf}

\input psfig.sty

\newcommand{\bq}{\begin{equation}}
\newcommand{\eq}{\end{equation}}
\newcommand{\bqn}{\begin{eqnarray}}
\newcommand{\eqn}{\end{eqnarray}}

\newcommand{\lb}{\label}

\def\be{\begin{equation}}
\def\ee{\end{equation}}
\def\lb{\label}

\def\gappr{\lower 3pt\hbox{$\buildrel > \over \sim\;$}}
\def\gappl{\lower 3pt\hbox{$\buildrel < \over \sim\;$}}
\def\limiter{\lower
7pt\hbox{$\buildrel{\textstyle\longrightarrow}\over{\scriptscriptstyle
~~s\rightarrow\infty~~}\;$}}
\def\ablim{\lower
9pt\hbox{$\buildrel{\textstyle\longrightarrow}\over{\scriptscriptstyle
~~~a\rightarrow b~~~}\;$}}
\def\x0lim{\lower
11pt\hbox{$\buildrel{\textstyle\longrightarrow}\over{\scriptscriptstyle
~~x^0\rightarrow-\infty~~}\;$}}
\def\xlim{\lower
8.5pt\hbox{$\buildrel{\textstyle\longrightarrow}\over{\scriptscriptstyle
~~x_\pm\rightarrow-\infty~~}\;$}}
\def\T0lim{\lower
11pt\hbox{$\buildrel{\textstyle\longrightarrow}\over{\scriptscriptstyle
~~T\rightarrow0~~}\;$}}
\def\Tlim{\lower
8.5pt\hbox{$\buildrel{\textstyle\longrightarrow}\over{\scriptscriptstyle~~T\rightarrow\infty~~}\;$}}
\def\Tmg1{\lower
8.5pt\hbox{$\buildrel{\textstyle\longrightarrow}\over{\scriptscriptstyle~~T>>1~~}\;$}}
\def\pdot{\raise 1.5pt\hbox{.}}

\def\im{\rm i}

\def\dal{\hbox{$\sqcup$\hbox to 0pt{\hss$\sqcap$}}}

        \def\im{\rm i}
         \def\ln{{\rm ln}}
         
         \def\dal{\hbox{$\sqcup$\hbox to 0pt{\hss$\sqcap$}}}

\def\mef{{M^*}}

\begin{document}


\title{Consistent nonadditive approach and nuclear equation of state}

\author{A. P. Santos$^{1}$} \email{alysonpaulo@dfte.ufrn.br}
\author{F. I. M. Pereira$^{2}$} \email{flavio@on.br}
\author{R. Silva$^{1,3}$} \email{raimundosilva@dfte.ufrn.br}
\author{J. S. Alcaniz$^{2}$} \email{alcaniz@on.br}

\affiliation{$^{1}$Departamento de F\'{\i}sica, Universidade Federal do Rio Grande
do Norte, 59072-970 Natal,  RN, Brazil}
\affiliation{$^{2}$Observat\'orio Nacional, Rua Gal. Jos\'e Cristino 77, 20921-400
Rio de Janeiro RJ, Brasil}
\affiliation{$^{3}$Universidade do Estado do Rio Grande do Norte, 59610-210,
Mossor\'o, RN, Brasil}


\begin{abstract}
By using a $q$-calculus, the Walecka many-body field theory was studied in the
context of the Tsallis framework.
 The most important aspect of the application of the $q$-calculus to the nonadditive
formulation of QHD-I is that it naturally emerges as a thermodynamically consistent
theory.

\end{abstract}

\maketitle

\section{Introducton}
\lb{intro}

 The properties of nuclear matter have been investigated through the relativistic
phenomenological approach introduced by Walecka {\it et al.} \cite{JDW,ChW,SW1}.
 This framework, also called quantum hadrodynamics (QHD-I), has been an important
approach
to  investigate the behavior of strong interactions at the hadronic energy scales.
 From the thermodynamical point of view, this model provides a consistent framework
for the
description of bulk static properties of strong interacting many-body nuclear systems.
 As a matter of fact, QHD-I is a strong-coupling renormalizable field theory of
nucleons
interacting via the exchange of scalar ($\sigma$) and vector ($\omega$) mesons
\cite{JDW,ChW,SW1}.
 Although being a consistent framework, some limitations of QHD-I have been
investigated
considering either a more complete theoretical study for zero temperature
\cite{chin,frning}
or by introducing the thermal field dynamics in hot nuclear matter \cite{heinning}.
 However, we are investigating the nonadditive effects on the QHD-I, as originally
developed in
\cite{JDW,ChW,SW1}.
 QHD-I has been used in the calculations of nuclear matter and finite nuclei (See
\cite{SW2} and
references therein), as well as in the astrophysics context to describe the
properties of nuclear
matter in compact stars
\cite{Gle,Menezes,Providencia,Santos,Grohmann,Dexheimer,Cavagnoli}.
 In this regard, some improvements and extensions of the model have been made to
make it more
suitable for neutron and proto-neutron stars calculations
\cite{Gle,Menezes,Providencia,Santos,Grohmann,Dexheimer,Cavagnoli}.

 On the other hand, the nonadditive statistics has received considerable attention,
both from
theoretical and observational viewpoints \cite{TsallisBook}.
 From the high energy point of view, the very first application was addressed to the
solar neutrino
problem by the use of the Tsallis framework to derive a distribution function for
the interior
plasma \cite{kaniadakis9697,Quarati97}.
Recently, by using the nonadditive statistics, the distribution of transverse
momenta in
high-energy collisions of proton-proton, and heavy nuclei (e.g., Pb-Pb and Au-Au)
have been object
of a lot of investigations \cite{hignergy1,hignergy2,hignergy3,hignergy4,hignergy5,hignergy6}.
 Actually, the $q$-distribution has been tested in high energy physics, and results
have presented very high quality fits of the transverse momentum distributions made
by the STAR \cite{Abelev} and PHENIX \cite{Adare} collaborations at RHIC and by the
ALICE \cite{Aamodt} and CMS \cite{Khachatryan} collaborations at LHC.
 More recently, an interesting discussion about the Renyi and Tsallis formulas has been investigated through a mathematical formulation in the context of ideal gas \cite{biro2013}. Specifically, Tsallis and Renyi entropies are related as $S_{Tsallis}= K S_{Renyi}=\sum_i p_i K(-\ln p_i)$, with $K(S)={1\over a} (e^{aS}-1)$ and $a=1/C_0$. $C_0$ is the positive heat capacity and $a$ is a real parameter. This approach has been successfully discussed in the context of black hole thermodynamics and quark-gluon plasma \cite{biro2013a,biro2013b}.

 The connection between the Tsallis statistics and the relativistic
nuclear equation of state (EoS) has also been discussed in the context of high energy physics (see, e.g., \cite{drago2004}).
 In particular, the EoS plays an important role in the determination of the structure and  evolution
of the proto-neutron stars \cite{drago07}.
 In this regard, the calculation of EoS was performed in the context of Tsallis framework and the main physical properties of proto-neutron stars as well as their astrophysical implications have been discussed in \cite{lavagno2011}.

 In the attempt to implement the Tsallis statistics on QHD-I, we have proposed a nonadditive
EoS through a straightforward generalization of Fermi-Dirac distribution
\cite{prc,pla}. This issue has been claimed, in a recent connection between the nonadditive framework and high
energy physics, with respect to the so-called thermodynamical consistency \cite{CMP,CW1,CW2}.
 In this concern, by considering the first and second laws of thermodynamics, the authors have
shown that such consistency is possible in the context of Tsallis framework \cite{CMP,CW1,CW2}.

 Until recently, the attempts to derive a nonadditive EoS for QHD-I, from first
principles of Quantum Statistical Mechanics (QSM), have appeared as an unthinkable task to be
achieved. 
 Fortunately, the developments on the so-called $q$-calculus has opened an important
possibility to overcome this difficulty showing that it is a powerful method that has
been frequently used to investigate many problems, e.g. connections with Quantum
Statistical Mechanics \cite{Rsilva2010,Nobre2011}.

 In this paper, we present a new approach for the thermodynamical
consistency of QHD-I. By following the $q$-calculus \cite{borges,nivanen}, we show that the new scheme
provides a calculation for the Walecka nuclear EoS based on first principles of the QSM.
 As a matter of fact,  instead of earlier calculations, we have found that
thermodynamical quantities describing the EoS, i.e., pressure, energy density and number
density depend explicitly on the nonadditive parameter $q$.
 The important result of the present work is that the thermodynamical
consistency naturally has emerged from first principles without any additional assumption.
 Moreover, from the mathematical point of view, both the pressure and energy density
expressions are equivalent to the ones that have been introduced through other approaches
\cite{lavagno1,lavagno2,lavagno3,lavagno4,lavagno5,lavagno6}.
 However, it must also be emphasized that, before the development of the
$q$-calculus, prescriptions about the way as the non-additive distribution functions
 $n_{\mathbf{k}}(T,\nu)$ and  $\bar{n}_{\mathbf{k}}(T,\nu)$ enter equations of pressure,
energy density, number density and so on, have been considered {\it a posteriori} by the
use of thermodynamical arguments to achieve the claimed consistency \cite{CMP,CW1,CW2}.
 Finally, we show that the $q$-calculus also provides a prescription for the scalar
density expression in terms of the powers of the non-additive distribution functions, 
$n^q_{\mathbf{k}}(T,\nu)$ and $\bar{n}^q_{\mathbf{k}}(T,\nu)$, for positive values of  $q$.

 The aim of the present work is twofold. First, our purpose is to use the $q$-calculus to solve, from first principles, the commented thermodynamical inconsistency in \cite{prc,pla}.
 Second, the simplification adopted here by the use of QHD-I\footnote{
For a more complete theoretical treatment (for zero temperature), see \cite{Friman88}. A discussion on the thermal field dynamics in hot nuclear matter
was presented \cite{Henning95}.} allows us to show, in a more
consistent way, that the $q$-calculus is general to be used in any quantum system.

 This paper is organized as follows.
 In Section \ref{qcalc}, we present the basic formalism of the QHD-I theory as well as its
connection with Tsallis framework through the $q$-calculus.
 The main results are discussed in Section III.
 Finally, we summarize our main conclusions in Section IV.

\section{Walecka equation of state in the q-calculus}
\lb{qcalc}

\subsection{QHD-I: Basic Equations}

 We now recall the main aspects of QHD-I.
 The Lagrangian is given by (for details, see \cite{SW1})
 \bqn\label{delagr}
{\cal L}&=&\bar{\psi }[(i\gamma _\mu (\partial ^\mu-g_\omega\omega^{\mu})-
(M-g_\sigma\sigma)]\psi
+ \frac 12(\partial _{\mu} \sigma \partial ^{\mu}\sigma-m_{\sigma}^{2} \sigma^{2})\\
&-&\frac {1}{4}\omega_{\mu \nu}
\omega^{\mu \nu}
 + \frac{1}{2}m_{\omega}^{2}\omega_{\mu} \omega^{\mu}~,
\eqn
where $\omega_{\mu\nu}$ is the field tensor
\bq\lb{Fmunu}
\omega_{\mu\nu}=\partial_\mu\omega_\nu-\partial_\nu\omega_\mu\;.
\eq

 The Euler-Lagrange equations for the fields and nucleons are given by
\bq\lb{sigma}
(\dal+m^2_\sigma)\;\sigma=g_\sigma\;\bar{\psi}\psi\;,
\eq
\bq\lb{omega}
\partial_\mu\;\omega^{\mu\nu}+m^2_\omega\;\omega^\nu=g_\omega\;\bar{\psi}\gamma^\nu\psi\;,
\eq
\bq\lb{dirac}
[\gamma^\mu(\im\partial_\mu-g_\omega\omega_\mu)-(M-g_\sigma\sigma)]\psi=0\;,
\eq
with the scalar source $\bar{\psi}\psi$ and the (conserved) baryon current
$B^\nu=\bar{\psi}\gamma^\nu\psi$.
 In the case of a system with uniform baryon and scalar densities, the meson fields
are replaced by the classical fields giving rise to the mean-field equations
\bq\lb{sigma2}
\sigma=\frac{g_\sigma}{m_\sigma^2}\langle\bar{\psi}\psi\rangle
\equiv\frac{g_\sigma}{m_\sigma^2}\;n_S\;,
\eq
\bq\lb{omega2}
\omega_0=\frac{g_\omega}{m_\omega^2}\langle\psi^\dagger\psi\rangle
\equiv\frac{g_\omega}{m_\omega^2}\;n_B\;\;\;
\eq
\bq\lb{dirac2}
[\gamma^\mu\im\partial_\mu-g_\omega\gamma^0\omega_0-(M-g_\sigma\sigma)]\psi=0\;,
\eq
with the scalar density $n_S=\langle\bar{\psi}\psi\rangle$ and baryon (vector) density
 $n_B=\langle\psi^\dagger\psi\rangle$; $m_\sigma$ and  $m_\omega$ are the corresponding
masses of the scalar and vector meson fields.
 The third term in Eq. (9) corresponds to the effective mass defined by
 \bq
 M^*=M-g_\sigma\sigma\;.
 \lb{em}
 \eq
 The QHD-I Hamiltonian and baryon number operator are given  by \cite{JDW,SW1}
\bq
\widehat{H}=\sum_{\bf{k}\lambda}E^*(k)(A^\dagger_{\bf{k}\lambda}A_{\bf{k}\lambda}+
B^\dagger_{\bf{k}\lambda}B_{\bf{k}\lambda})+g_\omega\omega_0\widehat
B-V\bigg(\frac{1}{2}m_\omega^2\omega_0^2-
\frac{1}{2}m_\sigma^2\sigma^2\bigg)\;,
\lb{H}
\eq

\bq
\widehat{B}=\sum_{\bf{k}\lambda}(A^\dagger_{\bf{k}\lambda}A_{\bf{k}\lambda}-
B^\dagger_{\bf{k}\lambda}B_{\bf{k}\lambda})\;,
\lb{B}
\eq
where $V$ stands for the volume of the system and $E^*(k)=\sqrt{k^2+\mef^2}$.
 In order to calculate the scalar density $n_S$ in the context of the $q$-calculus,
we need here an expression for the corresponding operator as done above for
$\widehat{B}$.
 To this end, by following the steps from Eq.~(3.28) to Eqs.~(3.41)~-~(3.44) in
\cite{SW1},
we obtain the expression for a scalar density operator, which we call
$\widehat{D}_S$, given by
\bqn
\widehat{D}_S&=&\frac{1}{V}\sum_{\bf{k}\lambda}\frac{M^*}{E^*(k)}
(A^\dagger_{\bf{k}\lambda}A_{\bf{k}\lambda}+
B^\dagger_{\bf{k}\lambda}B_{\bf{k}\lambda})\nonumber\\
&=&\frac{1}{V}\frac{\partial\widehat{H}}{\partial M^*}\;,
\lb{Dop}
\eqn
where the last line follows from Eq.~(\ref{H}).
 From Eqs.\;(\ref{H}) and (\ref{B}) it follows that
\bqn
\widehat H-\mu\widehat B &=&
\sum_{\bf{k}\lambda}\big[(E^*-\mu+g_0\omega_0)A^\dagger_{\bf{k}\lambda}A_{\bf{k}\lambda}+
(E^*+\mu-g_0\omega_0)B^\dagger_{\bf{k}\lambda}B_{\bf{k}\lambda}\big]\nonumber\\
&-&V\bigg(\frac{1}{2}m_\omega^2\omega_0^2-\frac{1}{2}m_\sigma^2\sigma^2\bigg)\;.
\lb{HB1}
\eqn
 In order to shorten the notation, we write
$\;\;\;\widehat N_{\bf{k}\lambda}=A^\dagger_{\bf{k}\lambda}A_{\bf{k}\lambda}$~,
$\;\;\;\widehat{\bar N}_{\bf{k}\lambda}=B^\dagger_{\bf{k}\lambda}B_{\bf{k}\lambda}$~,
$\;\;\;a_{\pm,\;\bf{k}\lambda}=E^*(k)\mp\nu$~,
where $\nu\equiv\mu-g_\omega\omega_0$ (which plays the role of an {\it effective
chemical potential} as can be seen bellow in Eq.~(\ref{nqk})) of the present paper, and
$b=V(m_\omega^2\omega_0^2-m_\sigma^2\sigma^2)/2$\;.
 So, Eq.~(\ref{HB1}) is reduced to
\bq
\widehat H-\mu\widehat B = \sum_{\bf{k}\lambda}\big(a_{+,\;\bf{k}\lambda}\widehat
N_{\bf{k}\lambda}+
a_{-,\;\bf{k}\lambda}\widehat{\bar N}_{\bf{k}\lambda}\big) - b\;,
\lb{HB2}
\eq
with the correspondence
$\widehat N_{\bf{k}\lambda}(\widehat{\bar N}_{\bf{k}\lambda})\rightarrow
a_{+,\;\bf{k}\lambda}(a_{-,\;\bf{k}\lambda})\rightarrow$ baryon(antibaryon).
  The partition function (for details, see \cite{FeW}, Chapter 2, pages 36-38), in the
Boltzmann-Gibbs statistical mechanics, is given by
\bqn
{\cal Z}_{BG} &=&\sum_{n_1,...\bar{n_1},...}\langle n_1,...\bar{n_1},...|\;
\exp\big[-\beta\sum_i\big(a_{+,\;i}\;\widehat{N}_i+a_{-,\;i}\;\widehat{\bar N}_i\big)+
\beta b\big]
\;|n_1,...\bar{n_1},...\rangle\nonumber\\
&=&\sum_{n_1,...\bar{n_1},...}\langle n_1,...\bar{n_1},...|\;
\exp\big[-\beta\sum_i\big(a_{+,\;i}\;n_i+a_{-,\;i}\;{\bar n}_i\big)+
\beta b\big]\;|n_1,...\bar{n_1},...\rangle\;.
\lb{ZG2}
\eqn
where $n_i(\bar{n}_i)$ are the baryon(antibaryon) occupation numbers in the state
$i\equiv\bf{k}\lambda$~, and $\beta=1/k_BT$.
 Notice that $\exp\big[-\beta\sum_i\big(a_{+,\;i}\;n_i+a_{-,\;i}\;{\bar n}_i\big)\big]$
is a c-number, which is a classical number (to distinguish it from a quantum observable which
is an operator).
 Now, our goal is to derive the pressure, energy density, and vector and scalar densities
within the nonadditive QSM.
 To this end, by using the $q$-calculus, we first obtain the nonadditive generalization of Eq.~(\ref{ZG2}).

\subsection{QHD-I and nonadditive framework}

 In order to introduce the nonadditive effects under the QHD-I, let us consider
this new possibility by investigating the issue from first principles.
 As well known, the temporal evolution of quantum system is given by \cite{FeW}
\be\label{eq1}
i\hbar\frac{\partial\Psi(r,t)}{\partial t}=\hat{H}\Psi(r,t).
\ee
The evolution operator is unitary and transform a state at instant $t_0$ in another
one at
$t$, i.e. $\Psi(r,t)=U(t,t_0)\Psi(r,t_0)$, with $U(t_0,t_0)=\hat{1}$.
By using this evolution operator, we can rewrite the expression~(\ref{eq1}) as
\be\label{eq2}
\left[i\hbar\frac{\partial U(t,t_0)}{\partial t}-\hat{H}U(t,t_0)\right]\Psi(r,t_0)=0,
\ee
and the solution is given by
\be\label{eq3}
 U(t,t_0)=\exp[-i\hat{H}(t-t_0)/\hbar].
\ee
 Now, we introduce nonadditive effects, by using an evolution operator in the context
of Tsallis framework.
 These effects have considered statistical correlations and non-linearity in the temporal
evolution of the system.
 In this regard, a generalization of expression~(\ref{eq3}) has been proposed in
\cite{turco}
to be of the form
\be\label{eq4}
 U_q(t,t_0)=\exp_q[-i\hat{H}(t-t_0)/\hbar],
\ee
where $\exp_q$ is the generalized exponential function.
 We here adopt the prescription given in \cite{TPM} by
\begin{eqnarray}\lb{TPM}
\exp_q (x) :=\left\{
\begin{array}{ccc}
[1+(q-1)x]^{\frac{1}{q-1}} & \mbox{if} & x>0 \\\nonumber
[1+(1-q)x]^{\frac{1}{1-q}} & \mbox{if} & x\leq0\;,
\end{array}
\right.
\end{eqnarray}
which better describes the nonadditive quantum systems (fermions and bosons).
By expanding $U_q(t,t_0)$, we obtain
\begin{eqnarray}\label{eq5}
 U_q(t,t_0)&=&
\hat{1}-\frac{-i}{\hbar}\hat{H}t+\sum_{n=2}^{\infty}\frac{1}{n!}\left[\frac{-i}{\hbar}\hat{H}U(t,t_0)\right]^n
\nonumber\\
&\times&(1-q)^{n-1}\prod_{k=1}^{n-1}\left(\frac{1}{1-q}-k\right),
\end{eqnarray}
with $U_q(t_0,t_0)=\hat{1}$.
 We may describe Eq.~(\ref{eq4}) in terms of the eigenstates of $\hat{H}$ with  $t_0=0$ as
\be\label{eq6}
 U_q(t)=\sum_{n}e_q^{-iE_nt/\hbar}|\Psi_n\rangle\langle\Psi_n|\;.
\ee
Introducing the substitution $t=-i\hbar\beta$~(Wick rotation)~\cite{cond} in the above
expression, we obtain
\be\label{eq7}
 U_q(-i\hbar\beta)=\sum_{n}e_q^{-\beta
E_n}|\Psi_n\rangle\langle\Psi_n|=Z_q(\beta)\;.
\ee
 By recognizing $\beta$ as inverse of $k_BT$, where $k_B$ is the Boltzmann constant and
$T$ is the absolute temperature of the system, we have $Z_q(\beta)$ as the generalized partition
function of the quantum system.
 Now, let us define the generalized density operator related to canonical ensemble
given
by
\be\label{eq8}
\hat{\rho}_q(\beta)=\frac{e_q^{-\beta\hat{H}}}{Z_q(\beta)}.
\ee
 Similarly, for the grand canonical ensemble, by taking the correspondence
$\hat{H}\rightarrow\hat{H}-\mu\hat{B}$, we have the respective density operator
\be\label{eq9}
 \hat{\rho}_q(\beta)=\frac{e_q^{-\beta(\hat{H}-\mu\hat{B})}}{\Xi_q}\;,
\ee
where
\begin{eqnarray}\label{eq10}
\Xi_q &=& \text{Tr}[e_q^{-\beta(\hat{H}-\mu\hat{B})}]\nonumber\\
&=& \sum_{n_1,n_2,...}\langle n_1,n_2,...| e_q^{-\beta(\hat{H}-\mu\hat{B})}
|n_1,n_2,...\rangle
\end{eqnarray}
is the generalized grand canonical partition function. 
 Let us mention that, until this point, the framework is completely general since one can introduce any physical system through of its Hamiltonian.

 At this point, by following the same steps to obtain the Boltzmann-Gibbs partition
function in Eq. (\ref{ZG2}), we use $q$-calculus \cite{borges,nivanen}.
 So, from Eqs. (\ref{HB2}) and (\ref{eq10}), we obtain the generalized QHD-I grand
canonical partition function
\begin{eqnarray}\label{eq11}
\Xi_q &=& \sum_{n_1,...,\bar{n}_1,...}\langle
n_1,...,\bar{n}_1,...|\exp_q[-\beta\sum_i(a_{+,i}\hat{N}_i+a_{-,i}\hat{\bar{N}}_i)+\beta
b] |n_1,...,\bar{n}_1,...\rangle\nonumber\\
      &=& \sum_{n_1,...,\bar{n}_1,...}\langle
n_1,...,\bar{n}_1,...|\exp_q[-\beta\sum_i(a_{+,i}n_i+a_{-,i}\bar{n}_i)+\beta
b] |n_1,...,\bar{n}_1,...\rangle\;.
\end{eqnarray}
Using orthogonality relations and properties of $q$-product~\cite{borges,nivanen}
\begin{equation}
\exp_q\left[\sum_ix_i\right]=\prod_{i}{_{\otimes_q}} e_q^{x_i}=e_q^{x_1}\otimes_q
e_q^{x_2}\otimes_q...
\end{equation}
we can rewrite~(\ref{eq11}) as
\bqn
\Xi_q &=& e_q^{\beta
b}\otimes_q\bigg[\sum_{n_1,...,\bar{n}_1,...}\bigg(\prod_{i}{_{\otimes_q}}
e_q^{-\beta a_{+,i}n_i}\bigg)\otimes_q\bigg(\prod_{i}{_{\otimes_q}} e_q^{-\beta
a_{-,i}\bar{n}_i}
\bigg)\bigg]\nonumber\\
 &=& e_q^{\beta b}\otimes_q \bigg[\sum_{n_1,...,\bar{n}_1,...}\left(e_q^{-\beta
a_{+,1}n_1}\otimes_q e_q^{-\beta a_{+,2}n_2}...\right)\otimes_q\left(e_q^{-\beta
a_{-,1}\bar{n}_1}\otimes_q e_q^{-\beta a_{-,2}\bar{n}_2}...\right) \bigg]\nonumber\\
 &=& e_q^{\beta b}\otimes_q \left[\left(1+e_q^{-\beta
a_{+,1}}\right)\otimes_q\left(1+e_q^{-\beta
a_{+,2}}\right)...\right]\otimes_q\left[\left(1+e_q^{-\beta
a_{-,1}}\right)\otimes_q\left(1+ e_q^{-\beta a_{-,2}}\right)...\right]
\eqn

Therefore, we obtain
\begin{equation}\label{eq13}
\Xi_q = e_q^{\beta b}\otimes_q\left[\prod_{i}{_{\otimes_q}} \left(1+e_q^{-\beta
a_{+,i}}\right)\otimes_q\left(1+ e_q^{-\beta a_{-,i}}\right)\right],
\end{equation}
or, by considering the original notation, we finally get the desired generalization
of the partition function for the QHD-I theory,
\be\label{eq14}
\Xi_q = e_q^{\beta
V(m^2_\omega\omega_0^2-m^2_\sigma\sigma^2)/2}\otimes_q\left[\prod_{i}{_{\otimes_q}}
\left(1+e_q^{-\beta(E^*-\nu)}\right)\otimes_q\left(1+
e_q^{-\beta(E^*+\nu)}\right)\right]\;.
\ee
 By taking the $q$-log of $\Xi_q$ and introducing the continuous limit, we have
\be\label{eq16}
\ln_q(\Xi_q) =\frac{V\gamma_N}{(2\pi)^3}\int d^3k\left\{
\ln_q\left[1+e_q^{-\beta(E^*-\nu)}\right] + \ln_q\left[1+
e_q^{-\beta(E^*+\nu)}\right]\right\}+\beta
V(m^2_\omega\omega_0^2-m^2_\sigma\sigma^2)/2\;.
\ee
 Here, $\gamma_N$ is the multiplicity factor ($\gamma_N=2$ for pure neutron matter and
$\gamma_N=4$ for nuclear matter).
 By using the known thermodynamical relations \cite{Kap}, after some algebra, the
pressure,
energy density and baryon number density are given, respectively, by
\be\label{eq17}
P=\frac{m^2_\omega}{2}\omega_0^2-\frac{m^2_\sigma}{2}\sigma^2+\frac{\gamma_N}{3(2\pi)^3}\int
d^3k \frac{k^2}{E^*(k)}[n^q_{\mathbf{k}}(T,\nu)+\bar{n}^q_{\mathbf{k}}(T,\nu)],
\ee
\be\label{eq18}
\epsilon=\frac{m^2_\omega}{2}\omega_0^2+\frac{m^2_\sigma}{2}\sigma^2+\frac{\gamma_N}{(2\pi)^3}\int
d^3k E^*(k)[n^q_{\mathbf{k}}(T,\nu)+\bar{n}^q_{\mathbf{k}}(T,\nu)],
\ee
\be\label{eq19}
n_B=\frac{\gamma_N}{(2\pi)^3}\int d^3k
[n^q_{\mathbf{k}}(T,\nu)-\bar{n}^q_{\mathbf{k}}(T,\nu)].
\ee
where
\begin{equation}
\label{nqk}
n_{\mathbf{k}}(T,\nu)=\frac{1}{{\rm e}_q^{\beta[E^*(k)-\nu]}+1}\;\;\;\;\;{\rm
and}\;\;\;\;\;
\bar{n}_{\mathbf{k}}(T,\nu)=\frac{1}{{\rm e}_q^{\beta[E^*(k)+\nu]}+1}\;,
\end{equation}
are the nonadditive distributions for baryons and anti-baryons.
 Note that, in the $q\rightarrow1$ limit, the standard Fermi-Dirac distributions
$n_{FD}(T,\nu)$ and $\bar{n}_{FD}(T,\nu)$ are recovered.
 Additionally, as physically expected, when $T\rightarrow0$ we obtain
$n_{\mathbf{k}}(T,\nu)\rightarrow n_{FD}(T,\nu)$ and $\bar{n}_{\mathbf{k}}(T,\nu)\rightarrow \bar{n}_{FD}(T,\nu)$.
 This amounts to saying that for studies of nuclear matter at zero temperature or neutron stars interiors
(where, in nuclear scale, $T \simeq 0$) we do not expect any nonadditive signature.
 On the other hand, in heavy ions collision experiments or in the interiors of protoneutron stars, with typical stellar temperatures of several tens of MeV (1 MeV$=1.1065\times10^{10}$ K), nonadditive effects can be important. 

 To calculate the scalar density given by $n_S=\langle \hat{D}_S\rangle$, we assume
that the nonadditive expectation value of the scalar density operator can be given by
\bqn
\langle\hat{D}_S\rangle &=&
\frac{\text{Tr}[\hat{D}_S\;e_q^{-\beta(\hat{H}-\mu\hat{B})}]}{\Xi_q},\nonumber\\
&=&\frac{1}{V}\frac{1}{\Xi_q}\frac{\partial}{\partial M^*}
\text{Tr}[\hat{H}e_q^{-\beta(\hat{H}-\mu\hat{B})}]
\lb{D}
\eqn
and, from Eq.~(\ref{Dop}), after some algebra, we obtain
\be\label{eq20}
n_S=\frac{\gamma_N}{(2\pi)^3} \int d^3k
\frac{M^*}{E^*(k)}[n^q_{\mathbf{k}}(T,\nu)+\bar{n}^q_{\mathbf{k}}(T,\nu)]\;.
\ee
 Notice that here $n_S$ is given in terms of $n^q_\mathbf{k}(T,\nu)$ and
$\bar{n}^q_\mathbf{k}(T,\nu)$.
 We here emphasize that this expression of $n_S$ has also emerged as a consequence of
the $q$-calculus.

 Finally, from (\ref{sigma2}), (\ref{em}) and (\ref{eq20}), we obtain the equation
\be
M^*=M-(\frac{g_\sigma}{m_\sigma})^2\frac{\gamma_N}{(2\pi)^3} \int d^3k
\frac{M^*}{E^*(k)}[n^q_{\mathbf{k}}(T,\nu)+\bar{n}^q_{\mathbf{k}}(T,\nu)]\;,
\label{sceq}
\ee
from which the effective mass $M^*$ is determined self-consistently.

Additionally, we also use for the coupling constants the values given in reference
\cite{JDW,ChW,SW1}, namely\footnote{For the purpose of the present work, the values
given in Eq. (\ref{cpcts}) suffices to investigate the effects of the nonadditivity
in neutron and nuclear matter.
 Variations of the coupling constants, within the acceptable values given in current
literature, do not qualitatively affect the conclusions.},

\begin{equation}
\label{cpcts}
\bigg(\frac{g_\sigma}{m_\sigma}\bigg)^2=11.798~{\rm fm^2}~{\rm~and}~
\bigg(\frac{g_\omega}{m_\omega}\bigg)^2=8.653~{\rm fm^2}~,
\end{equation}
which are fixed to give (at $T=0$) the bind energy $E_{\rm bind}=-15.75$ MeV and $k_F=1.42$
$\rm{fm}^{-1}$.
 Other important nuclear quantities such as, for example, the symmetry energy coefficient
and compressibility, at saturation density, are calculated at zero temperature
(for a review, see \cite{Bla}).
 As commented above, at $T=0$ the effects of nonadditivity do not appear in these quantities.
 In particular, the problem of the high value of the compressibility predicted by QHD-I, with
respect to the empirical one (as shown in Table II of Ref. \cite{SW1}), is due to the limitations
of the theory.
 Consequently, in the present work, the predictions of QHD-I at $T=0$ remain the same as in its
original form, independently of the value of the $q$ parameter.

\section{Results}
\lb{res}

\begin{figure*}[th]
\centerline{
\psfig{figure=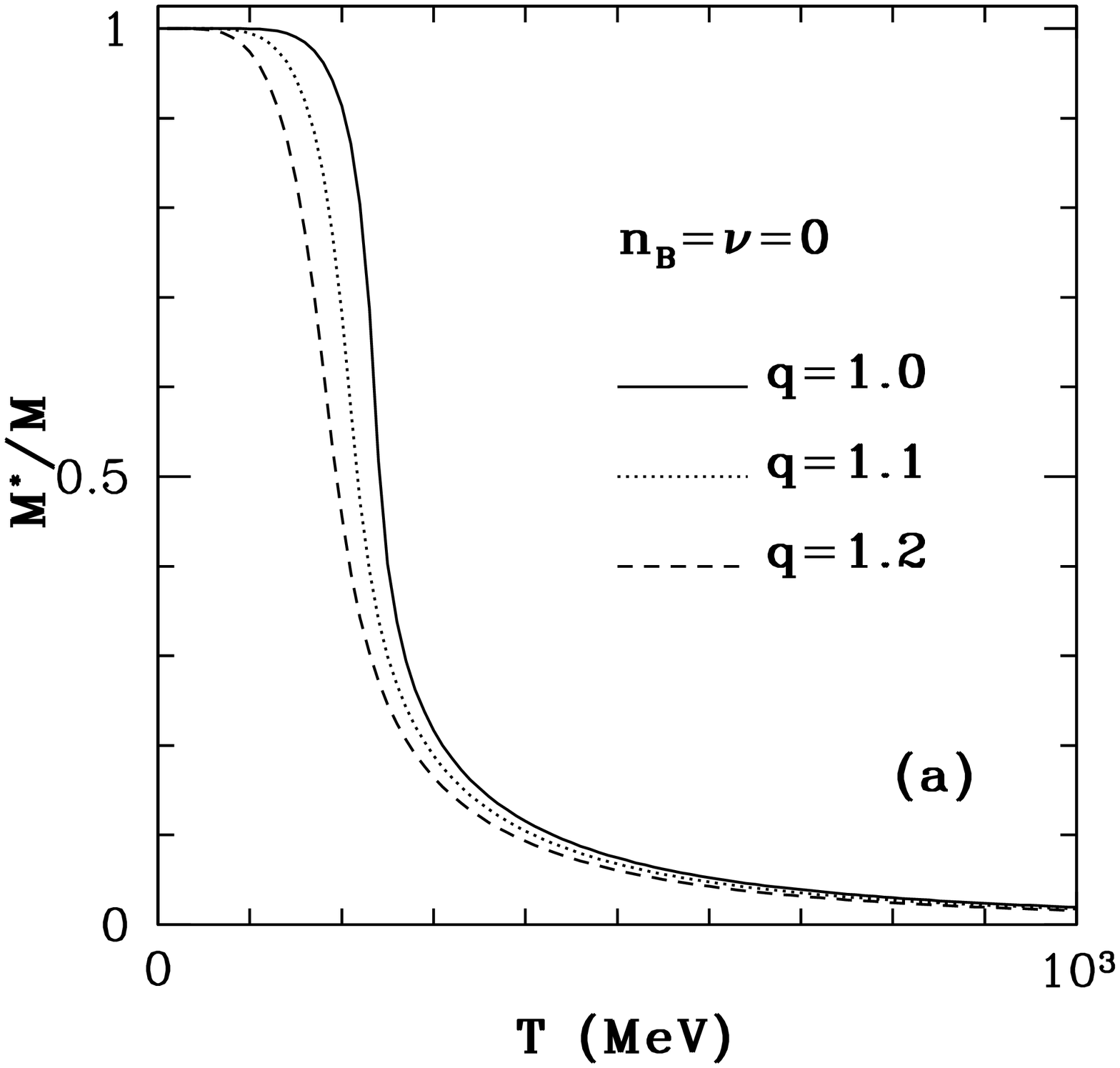,width=3.2truein,height=2.25truein}
\psfig{figure=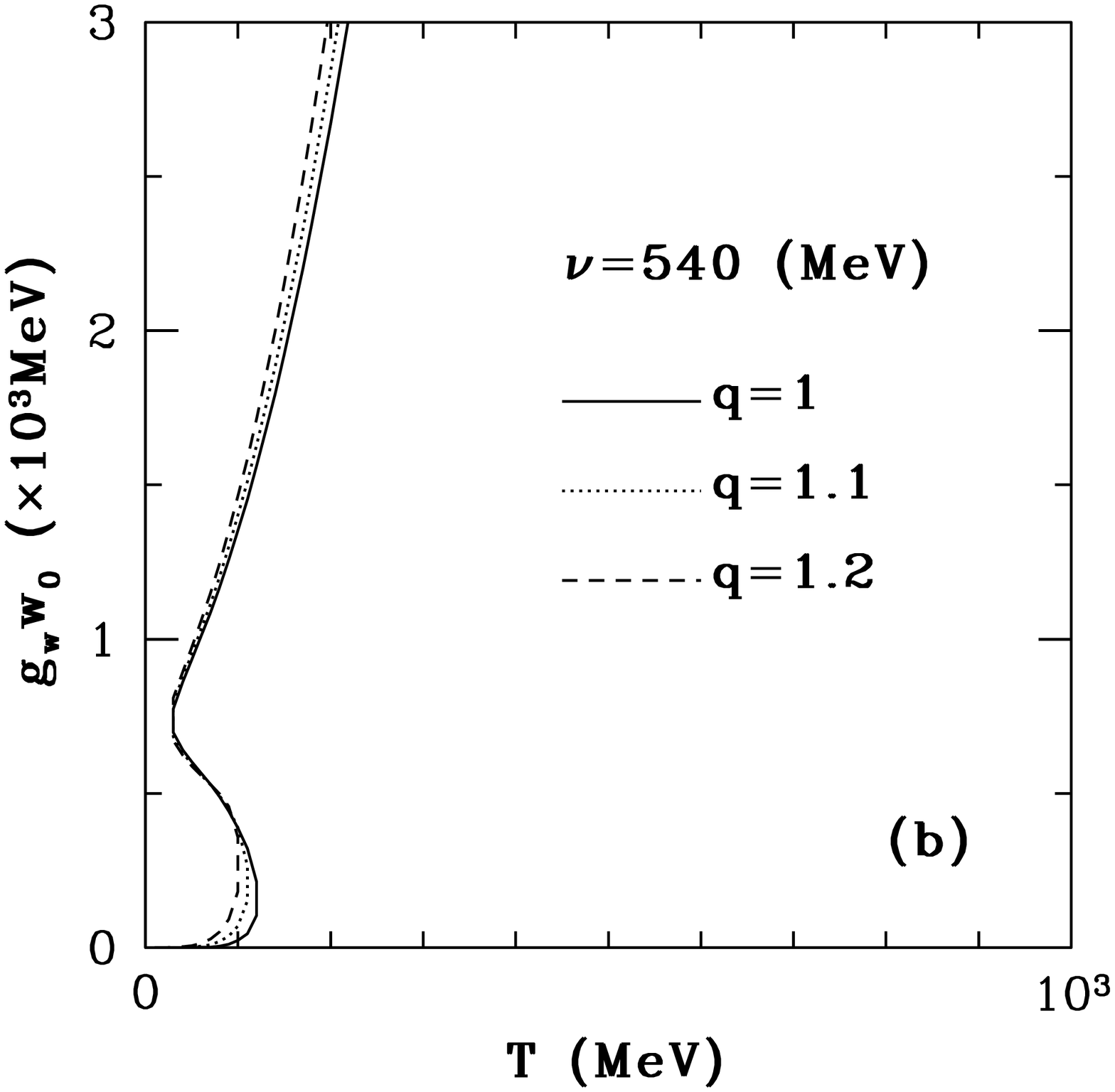,width=3.2truein,height=2.25truein}
\hskip .5in}
\caption{For pure neutron matter ($\gamma_N=2$):
 (a) the self-consistent nucleon mass at zero baryon density
as function of temperature for different values of the parameter $q$.
 (b) As in (a), but for the vector meson field at nonzero baryon
density for the given value of $\nu$.}
\label{mgw0}
\end{figure*}

 As done in \cite{prc,pla}, in the present paper, we have calculated the important
quantities that govern the behaviors of pure neutron matter ($\gamma=2$) and nuclear
matter ($\gamma=4$) for several values of $T$ and $q$.
 Generally speaking, the results are qualitatively similar to those in \cite{prc,pla},
except for some cases which we comment here.
 More specifically, with respect to the cited previous works, the main aspect of the
calculation by using the powers of the non-additive distribution functions, 
 $n^q_{\mathbf{k}}(T,\nu)$ and $\bar{n}^q_{\mathbf{k}}(T,\nu)$,
instead of $n_{\mathbf{k}}(T,\nu)$ and $\bar{n}_{\mathbf{k}}(T,\nu)$, is that the numerical
results (the curves) for the scalar and vector meson fields, pressure, energy density,
and so on, become closer to the corresponding ones for $q=1$.
 As an example, we show in Fig. \ref{mgw0}, for pure neutron matter, the effective
nucleon mass (in panel (a)) and the vector meson field (in panel (b)) as functions
of $T$, at the given values of $\nu$ and $q$.
  It is evident the similarity of the present results with respect to the ones in Refs.~\cite{prc,pla}.
 So, we here do not show all the corresponding figures, except for a few cases which
present some differences with respect to the ones in \cite{prc,pla}.

\subsection{The phase structure at non-zero baryon density}
\lb{phsnzbd}

 Of particular interest is the behavior of the EoS of the nuclear matter ($\gamma=4$)
around the critical point, where the liquid-gas phase transition changes from first to second
order.
 This phase transition is of the Van der Waals type.
 As is well known, at temperatures lower than the critical temperature $T_c$, the isothermals
of the $P\times\epsilon$ plot present three characteristics regions (for more detailed descriptions,
see \cite{Kap} and \cite{Pas}).
 The first one, at low energy densities, corresponds to a gas phase of (evaporated) nucleons.
 The second one, at intermediate energy densities, is a mixed phase made of coexisting nucleon gas and
nucleon liquid (formed by clumps or droplets of nucleons).
 Finally, the third region, at high energy densities, corresponds to the pure nucleon liquid.
 In the mixed phase, the $P\times\epsilon$ plot oscillates between a maximum and a minimum
(with a pronounced dip, in some cases), being usually treated via Maxwell construction.
 At $T<T_c$, the liquid-gas phase transitions are of first order.
 As $T$ increases, the intermediate region of the mixed phase shrinks, vanishing at $T=T_c$.
 In this case, the $P\times\epsilon$ plot has a turning point in which the system passes
(along the isotherm) from the gas phase to the liquid phase, characterizing a second order phase transition.
 For $T>T_c$, the system is in the gas phase \cite{Pas}.
 In nonadditive statistical mechanics, these features are dependent not only on the values of the coupling
constants of the QHD-I theory, but also on the value of the $q$ parameter.
 Moreover, the EoS becomes stiffer for increasing values of $q$.

\begin{figure*}[bh]
\centerline{
\psfig{figure=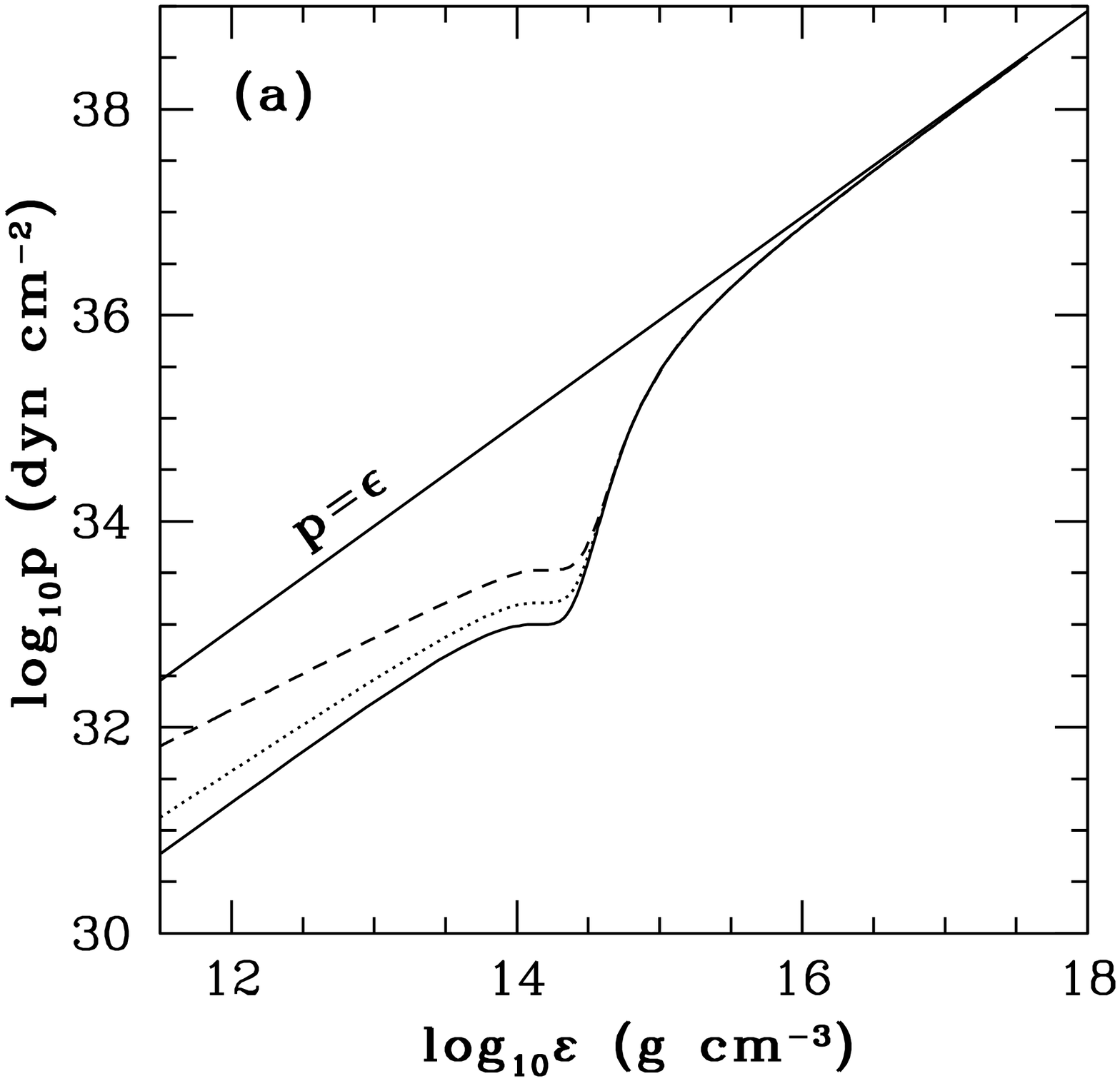,width=3.2truein,height=2.25truein}
\psfig{figure=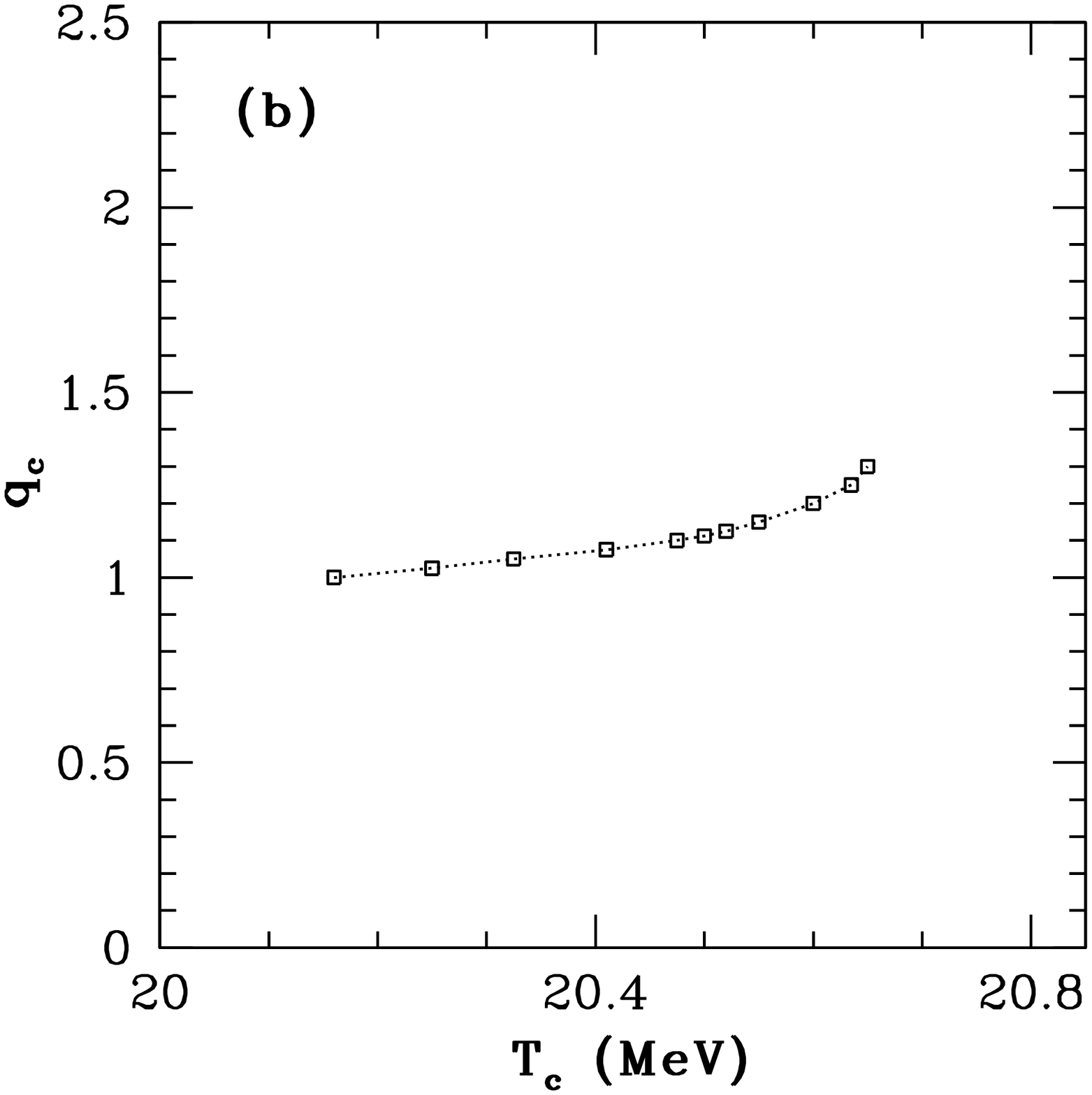,width=3.2truein,height=2.25truein}
\hskip .5in}
\caption{For nuclear matter ($ \gamma_N=4$):
 (a) isotherms for three arbitrarily chosen values of the critical
temperature $T_c$ and the corresponding parameter $q_c$ for
$T_c=20.15$ MeV and $q_c=1$ (solid);
$T_c=20.475$ MeV and $q_c=1.1$ (dots);
$T_c=20.635$ MeV and $q_c=1.25$ (dash).
 (b) The critical parameter $q_c$ as function of the critical temperature.}
\label{qctc}
\end{figure*}

 The dip observed for $q=1$, that characterizes the first order phase transition at
temperatures lower than the critical one, can be reduced if we take values of
$q>1$, suggesting that Maxwell construction can be canceled by properly choosing the
values of $q$.
 However, this is possible only for a restricted range of $T$ and $q$.
 A detailed analysis to obtain the critical temperatures and corresponding nonadditive
parameters have shown that for $q=1$, the critical temperature is $T_c=20.16$ MeV,
illustrated by the lower solid curve in panel (a) of Fig. \ref{qctc}.
 Below this temperature, Maxwell construction is not eliminated.
 The upper curve corresponds to $T_c=20.65$ MeV and $q_c=1.3$ (which is practically
in the convergence limit of the integrals in Eqs. (\ref{eq17})-(\ref{eq19}) and
(\ref{eq20}).
 In this range of temperatures, we have calculated several values of $T_c$ and $q_c$,
plotted in panel (b).
 We remark that the turning points of the isotherms in panel (a) are slightly shifted
to right, corresponding to different (increasing) values of the energy density (or the
baryon number density).
 Similarly, in panel (b), the baryon density increases along the curve from left to right.
 The nonlinearity of the points clearly show the influence of the $q$ power in
$n^q_{\mathbf{k}}(T,\nu)$ and $\bar{n}^q_{\mathbf{k}}(T,\nu)$, when compared with the (inconsistent)
use of $n_{\mathbf{k}}(T,\nu)$ and $\bar{n}_{\mathbf{k}}(T,\nu)$ in \cite{prc,pla}.

\subsection{The phase structure at zero baryon density}
\lb{phszbd}

 By considering the same arguments of \cite{TSP} and \cite{pla}, we have explored
the nonadditive phase structure of QHD-I at vanishing baryon density, for the value
of the coupling constant $C_S^2=365$, where $C_S^2=(g_\sigma/m_\sigma)^2 M^2\;$
\cite{TSP}.
 At $n_B=\nu=0$, the terms with the baryon density do not appear in
Eqs. (\ref{eq17}) and (\ref{eq18}).

\begin{figure*}[th]
\centerline{
\psfig{figure=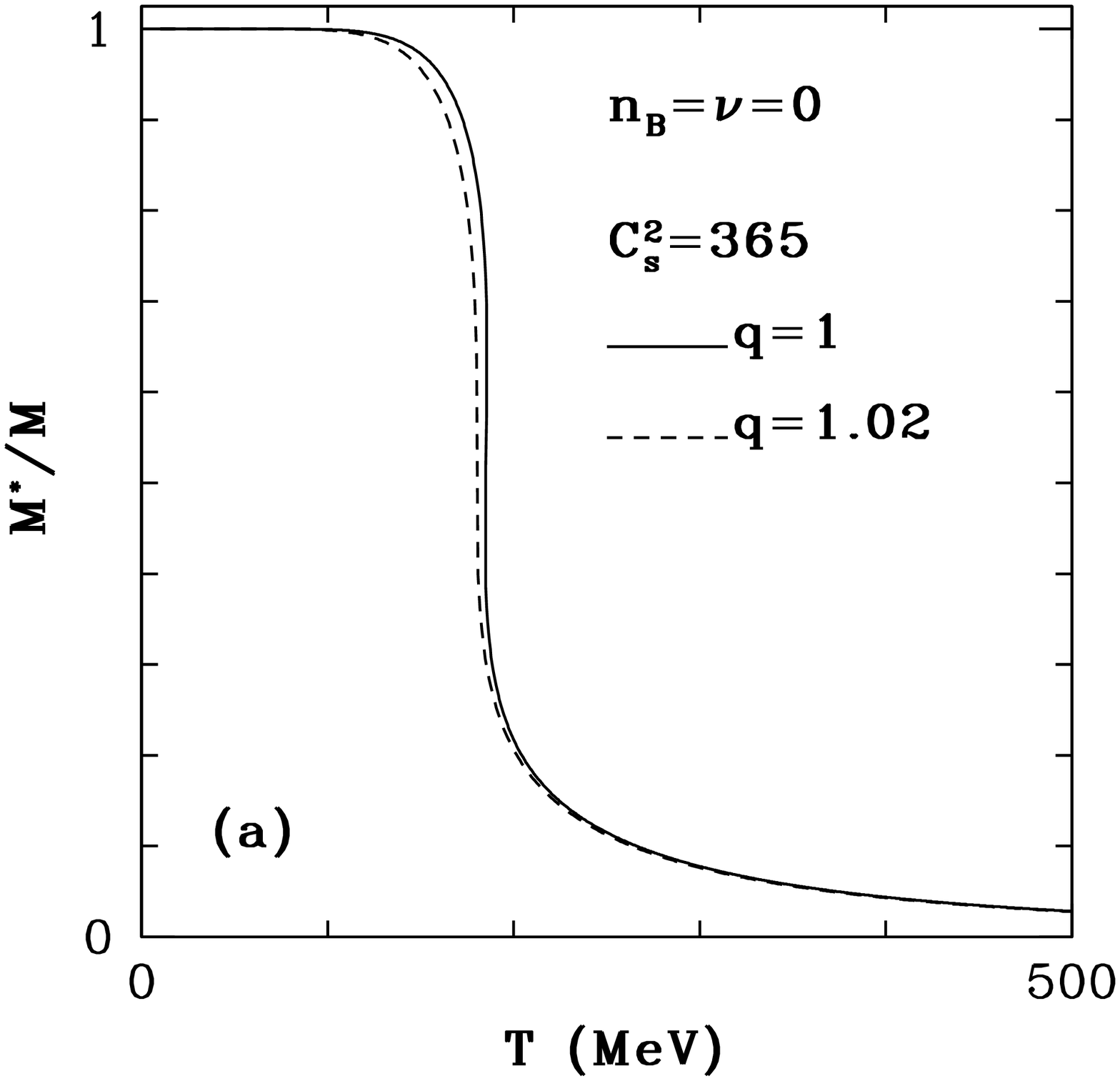,width=3.2truein,height=2.25truein}
\psfig{figure=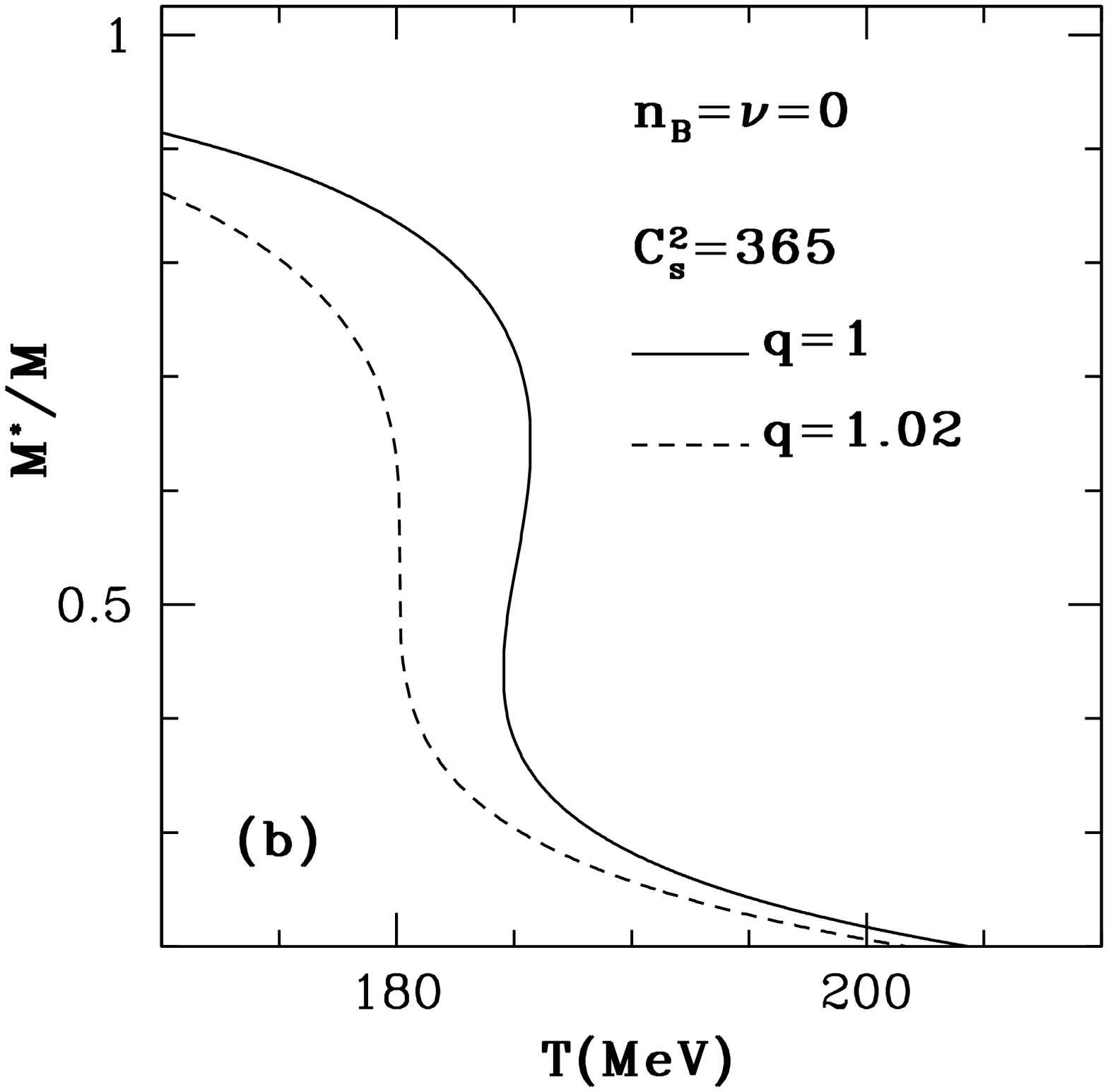,width=3.2truein,height=2.25truein}
\hskip .5in}
\caption{For nuclear matter ($ \gamma_N=4$):
(a) the self-consistent nucleon mass at zero baryon density as function
of temperature at the given values of $C_s^2$ and $q$.
 (b) As in (a), but with a stretched temperature region around the point of
phase transition.}
\label{qmn12}
\end{figure*}

 In Fig.~\ref{qmn12}, the panel (a) shows for $q=1$ a sudden drop in $M^*$.
 In reality, the detailed curve, in panel (b), shows that around $T\sim185\;{\rm MeV}$
the self consistent solution of Eq.~(\ref{sceq}) is triple valued.
 The pressure and energy density are also triple valued, with abrupt
rises at the same temperature, characterizing the first order phase transition.
 As a consequence of the sudden drop making the effective mass very small at high
temperatures ($T\gappr185~$MeV), the system decouples to a nearly free-massless nucleon gas.
Here, this depends on the values of $C_s^2$ and $q$.
 As pointed out in \cite{TSP}, this is analogous to the expected chiral phase transition
in high-temperature QCD.  However, in Walecka theory, there is no liberation of the internal
constituents of the nucleons. Thus, the phase transition in QHD theory can not be interpreted
as a baryon-quark matter one.

 On the other hand, interesting features are obtained for the specific heat.
 Differently from the treatment discussed in  Ref.~\cite{TSP}, the mathematical
structure of the self-consistency equation in our approach is not simple, so that the
calculation must also be done numerically.
 We observe that the specific heat calculated from Eq.~(\ref{eq18}) is linear in
$dM^*/dT$.
 By writing
\begin{equation}\label{Ch}
C_H=\frac{d\varepsilon}{dT}=\frac{d\varepsilon}{dM^*}\frac{dM^*}{dT}\;,
\end{equation}
we can see from Eq.~(\ref{sceq}) that
\bqn\label{dMdT1}
\frac{dM^*}{dT}&=&\frac{-2C_{M^*}\frac{M^*}{T}\int_0^\infty\frac{2k^2+M^{*2}}{E^*(k)}\;
n^q_{\mathbf{k}}(T,\nu)\;dk}{1+2\;C_{M^*}\int_0^\infty\frac{ k^2-M^{*2}}{E^*(k)}\;
n^q_{\mathbf{k}}(T,\nu)\; dk}\;\\\nonumber
&=&\frac{-2C_{M^*}\frac{{M^*}^2}{T}\int_0^\infty\frac{2k^2+M^{*2}}{E^*(k)}\;
n^q_{\mathbf{k}}(T,\nu)\;dk}{M-2\;C_{M^*}{M^*}^3\int_0^\infty n^q_{\mathbf{k}}(T,\nu)
\frac{dk}{E^*(k)}}\;,
\eqn
where
$$C_{M^*}=(g_\sigma/m_\sigma)^2\gamma_N/\pi^2\equiv C_S^2/M^2.$$

\begin{figure*}[tbh]
\centerline{
\psfig{figure=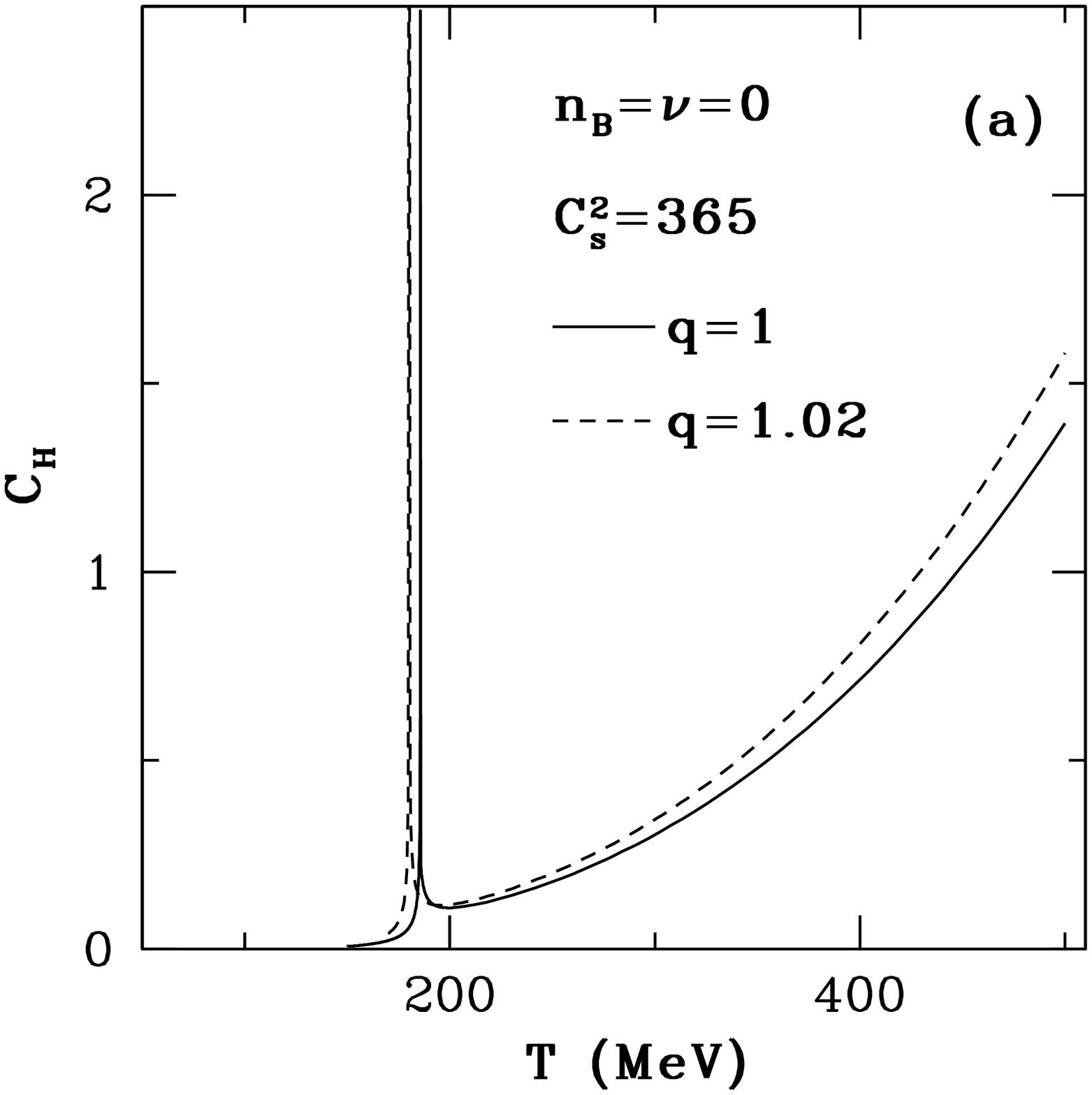,width=3.2truein,height=2.25truein}
\psfig{figure=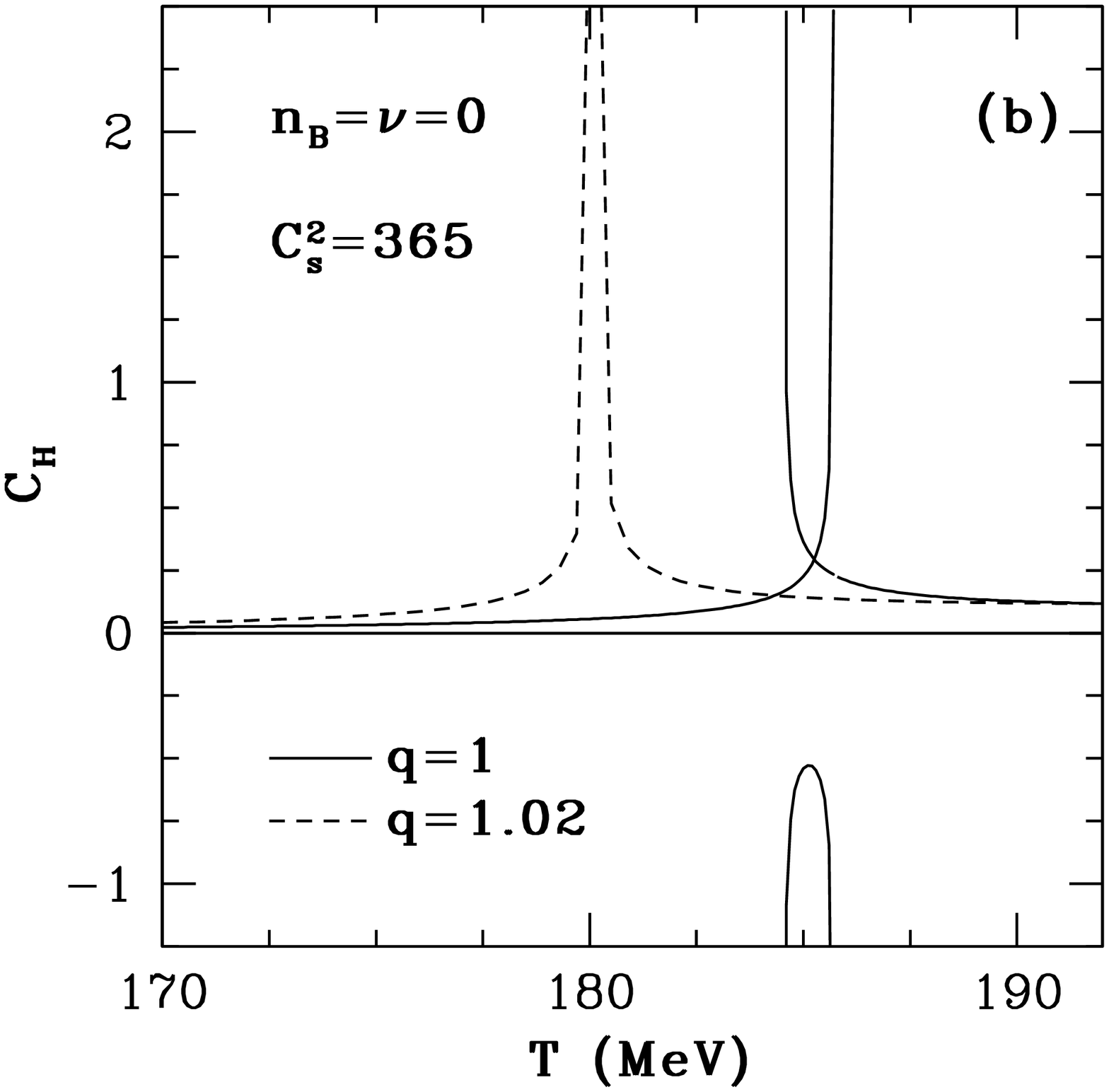,width=3.2truein,height=2.25truein}
\hskip .5in}
\caption{(a) The specific heat of nuclear matter ($\gamma=4$) divided by the corresponding
$q=1$ Stefan-Boltzmann limit ($C_H^{SB}=\gamma_N(7\pi^2/30)k_B^4T^3$) as function
of temperature for the given value of $C_S^2$ and two different values of the parameter $q$.
 (b) The same as in (a), but in the stretched region around the phase transition point.
 For $q=1$ and $q=1.02$ the phase transitions are, respectively, of the first and
second order.}
\label{pcv12}
\end{figure*}

 The singularities of $dM^*/dT$ lie on the curve determined by the vanishing of the
denominator.
 We observe that, when there is a sudden fall in $M^*$, a peak arises in the specific
heat  around $T\sim185\;{\rm MeV}$, as shown in panel (a) of Fig. \ref{pcv12}.
 Panel (b) shows the complicated structure of the specific heat (with a negative part)
characterizing a first order phase transition.
 For $q=1.02$, $C_H$ is single valued and presents a smooth behavior with a
peak around $T\sim180\;{\rm MeV}$ (instead of, say, $T\sim175\;{\rm MeV}$ in \cite{pla}).
 We also observe that, at zero baryon density,  Eq.~(\ref{eq19}) gives an equal
number of nucleons and anti-nucleons, so its ratio is $N/\bar{N}=1$ at nonzero temperatures,
independently of the values of the $q$ parameter.

\section{Final remarks}

 The thermodynamics consistency from the nonadditive framework in high energy physics is
fundamental in the construction of a coherent approach.
 In this paper, we have explored the mean field theory of Walecka (QHD) \cite{JDW,ChW,SW1,SW2}
in the context of the non-additive framework without additional requirements.
 The treatment discussed here differs of the previous results investigated in \cite{prc,pla},
i.e., a simple substitution of Fermi-Dirac distribution by a nonadditive quantum distribution
was considered to calculate the Walecka nuclear EoS.

 By considering the so-called $q$-calculus \cite{borges,nivanen}, we have introduced a generalized
thermodynamical formulation based only on the first principles of quantum statistical mechanics.
 The thermodynamics variables describing the nuclear matter, i.e, pressure, energy density,
number density and scalar density depend on the nonadditive parameter being given by expressions
(\ref{eq17})-(\ref{eq19}), (\ref{eq20}) and (\ref{sceq}).

 We have investigated the phase structure of nuclear matter at high temperature, by considering
both zero and non-zero baryon density, as well as a pure neutron matter ($\gamma = 2$) and a
nuclear matter ($\gamma = 4$).
 As discussed earlier, differently of previous results based on the inconsistent thermodynamical
description \cite{prc,pla}, we have shown that the nonlinearity from expressions
(\ref{eq17})-(\ref{eq19}) has not eliminated the Maxwell constructions (except for a very restricted
range of temperature).
 By following similar arguments of \cite{pla,TSP}, we have calculated numerically the specific heat,
being that the order of phase transitions also depends of nonadditive parameter $q$.

 The most important aspect of the present work rests on the $q$-calculus with which we have
shown that, from first principles, a thermodynamically consistent calculation of the QHD-I
equation of state is possible.
 In spite of limitations presented by QHD-I, we have used it as an example to show the effects
of nonadditivity on the nuclear EoS.
 We here have also shown, for the first time, the importance of the $q$-calculus, due to its
generality for application in quantum systems.
 Mathematically speaking, the present use of $q$-calculus serve to show that its generalization to extended
forms of QHD-I is straightforward.

 Although we have applied the $q$-calculus on the QHD-I theory, the formalism is in principle
general for any physical system described by a partition function like that in Eq. (\ref{eq10}).
\vspace{.5truecm}

\noindent{\bf Acknowledgements}

The authors (APS, RS and JSA) thank INCT-INEspa\c co, MCT, CNPq and FAPERJ
for the grants under which this work was carried out.

\newpage







\vspace{2.truecm}

\newpage

\vspace{2cm}

\vspace{2.truecm}

\end{document}